\documentclass[prl,twocolumn,preprintnumbers,floatfix]{revtex4}
\usepackage{graphicx}
\usepackage{verbatim}
\begin{document}

\title{
Model of fluorescence intermittency of single colloidal
    semiconductor quantum dots using multiple recombination centers}
\author{ Pavel~A.~Frantsuzov, S\'andor Volk\'an-Kacs\'o and Bolizs\'ar Jank\'o}
\affiliation{Department of Physics,
University of Notre Dame, Notre Dame, IN 46556, USA}

\begin{abstract}

We present a new physical model resolving a long-standing mystery
of the power-law distributions of the blinking times in single colloidal quantum dot fluorescence. The model
considers the non-radiative relaxation of the exciton through multiple
recombination centers. Each center is allowed to switch between two quasi-stationary
states. We point out that the conventional threshold analysis method used to extract the
 exponents of the distributions for the on-times and off-times
has a serious flaw: The qualitative properties of the distributions strongly depend
on the threshold value chosen for separating the on and off states.
Our new model explains naturally this threshold dependence, as well as other
 key experimental features of the single quantum dot fluorescence trajectories, such as the
power-law power spectrum (1/f noise).

\end{abstract}
\date{\today}
\maketitle

Substantial progress has been made recently in the study of
long range correlations in the fluctuations of the emission intensity (blinking) in
single colloidal semiconductor nanocrystals (QD)
\cite{BrusNature96,KunoJCP00,BawendiPRB01,MulvaneyPCCP06,OrritCOCIS07},
nanorods \cite{CrouchJPCB06}, nanowires\cite{KunoAM05} and some
organic molecules\cite{HoogenboomCPC07}.  By introducing
an intensity threshold level to separate bright (on)
and dark (off) states,  Kuno et al. \cite{KunoJCP00} found that the on-
and off-time distributions in QDs exhibit a spectacular power-law
dependence over 5-6 orders of magnitude in time.
\begin{equation}
p_{\mbox{on/off}}(t)\sim t^{-m}
\label{eq:PowerLaw}
\end{equation}
As discovered later by Shimizu et
al. \cite{BawendiPRB01}, the power-law on-time distribution is cut off at times ranging from a few seconds to 100 \ s, depending on the dot and its environment. During the past eight years or so the truncated power-law form of the blinking on-time distributions
was confirmed by many experimental groups (see \cite{MulvaneyPCCP06,OrritCOCIS07,BarkaiPT09,NesbittNL09} and references therein), but its microscopic origin remains a mystery.
Remarkably,  there are no "set" values for the on-time and off-time exponents. They are scattered in the region from 1.2 to 2.0.
Similar on- and off-time distributions were found recently for the other blinking systems mentioned above: semiconductor nanorods(NRs)\cite{CrouchJPCB06,DrndicNL08}, nanowires (NWs)\cite{KunoAM05} and organic dyes \cite{HoogenboomCPC07}.
The generality of the phenomenon is rather intriguing. We argued
that there must be a common underlying mechanism responsible for
the long time correlated fluorescence intermittency detected in all these systems\cite{NaturePhys08}. Most theoretical explanations of the QD
blinking \cite{KunoJCP00,BawendiPRB01,OrritPRB02,BarkaiJCP04,TangPRL05}
are based on the Efros/Rosen charging mechanism \cite{EfrosPRL97}.
The mechanism attributes on- and off- states to a neutral and a charged QD,
respectively. The light-induced electronic excitation in the charged
QD is quenched by a fast Auger recombination process. A
number of experimental results indicate, however, that there are no
unique bright  (on) and dark  (off) states of the QD, but a continuous
set  of emission intensities \cite{MewsPRL02,FisherJPCB04,YangNL06}.
One can therefore suggest an alternative mechanism of the blinking, assuming slow
fluctuations in the non-radiative recombination rate of the excited state   \cite{FrantsuzovPRB05,BarnesJPCB06,BarbaraCP07}.

In order to gain further insight into the possible blinking mechanism, we performed an extensive analysis of the on- and off- distributions of actual
single QD fluorescence trajectories.
Our procedure is different from the conventional ones, as we applied the Maximum Likelihood Estimator (MLE)  method to find the best Gamma-distribution $p(t)\sim t^{-m}\exp(-t/T)$  fit for the set of on and off durations. The MLE approach \cite{HoogenboomJCP06} gives an unbiased estimation for the parameters of the power-law distribution with minimal statistical error. These properties are crucial and allow for the investigation of a {\it single} trajectory.  Our approach, in contrast with  procedure used by Hoogenboom et al. \cite{HoogenboomJCP06}, allows us to find optimal values for  not only for $m$, but for the truncation time $T$ as well.
The fluorescence trajectories we investigated
were obtained by Protasenko and Kuno and have already been analyzed by others \cite{BarkaiACP06,GrigoliniJCP05}.

\begin{figure}[ht]
\includegraphics[width=3 in,clip]{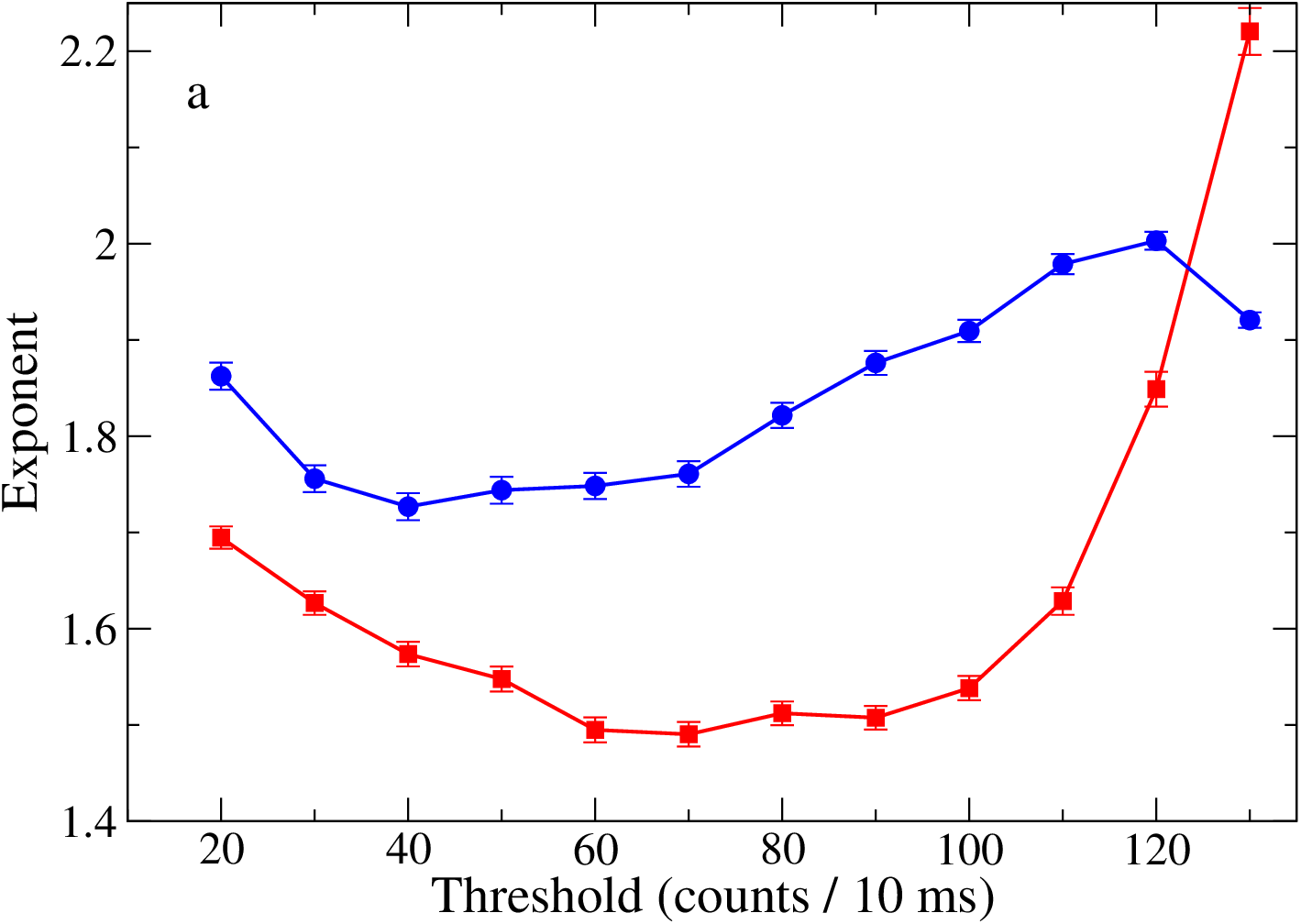}
\hfill
\includegraphics[width=3 in,clip]{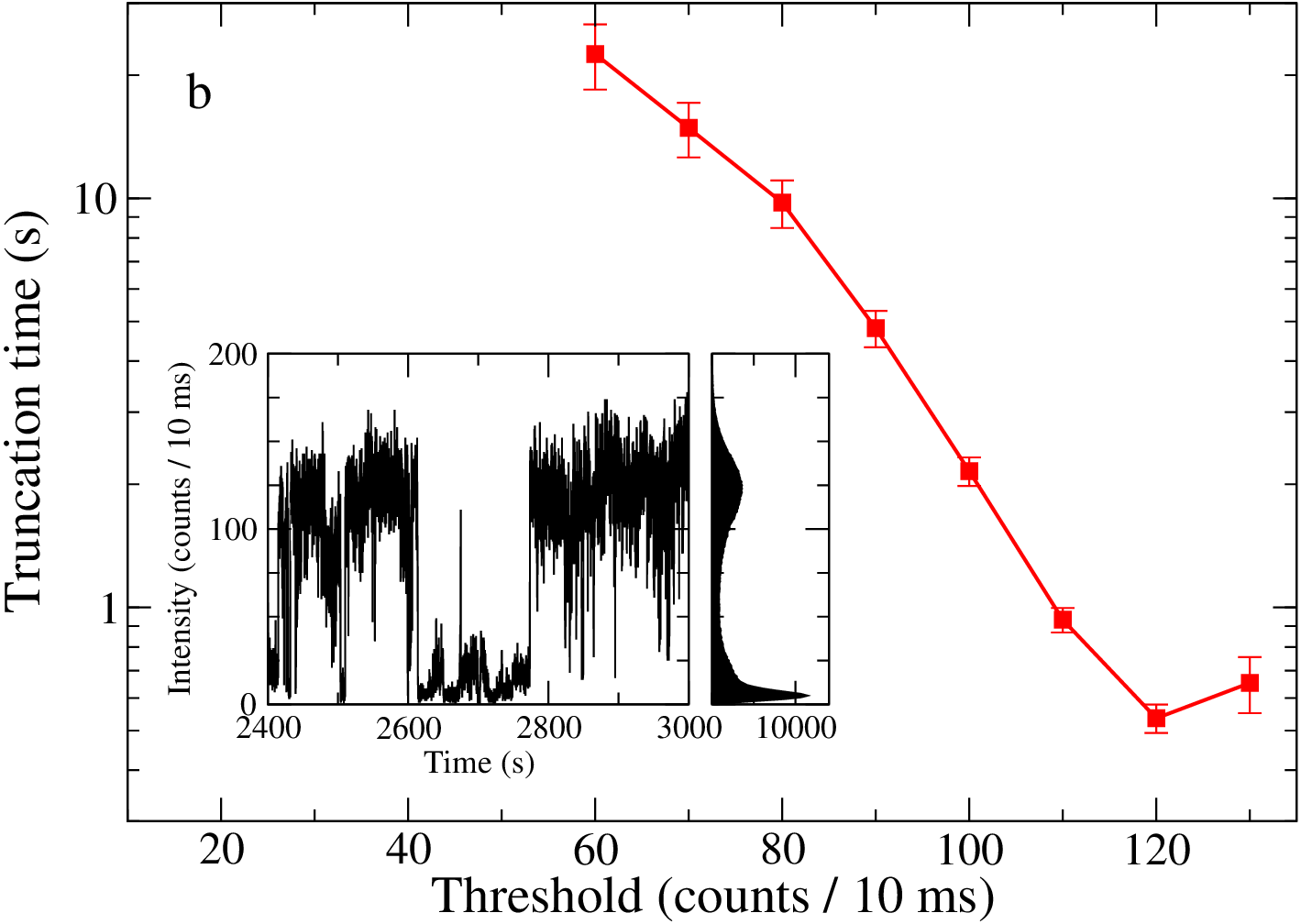}
\caption{(color online) The threshold dependence of the on-time (red squares) and off-time (blue circles) distribution exponents  (a) and on-time distribution truncation time (b)
obtained from the experimentally measured single QD fluorescence trajectory. Error bars show the standard deviation values. Insert b: a part of the trajectory and an intensity histogram.    }
\label{fig:trajectory}
\end{figure}

Our fitting procedure is performed repeatedly for a number of threshold values for each trajectory.
 In all  the cases the off-time distribution truncation time is found too long to be detected.
 Also, the threshold dependence of the distributions was all but ignored until now. The only exception is the recent observation made on nanorods (not QDs) by Drndic
 and her coworkers \cite{DrndicNL08}. In any case, the fundamental nature of this dependence was not revealed until now. An example of the  threshold dependency  of
 the power law exponent (slope on log-scale) and on- truncation time  for a singe QD trajectory is presented in Fig \ref{fig:trajectory}.
 While we investigated a large number of trajectories, we have deliberately chosen for this paper one with clearly visible telegraph noise-like features and well-defined on and off maxima in
 the intensity histogram [see inset in Fig. 1(b)]. As it is evident from Fig \ref{fig:trajectory}, even for this apparently ideal case, the distribution parameters are strongly threshold dependent.
 While the majority of the analyzed trajectories are not like telegraph noise, we mention that {\it all} show similar threshold dependence. The on-time truncation time decreases monotonically with
 increasing of the threshold. This trend is the same for most  single QD fluorescence trajectories we analyzed. The scaling of the slope as a function of the threshold is more complicated.
 The exponent of the off-time distribution shows several extrema, whereas the on-time exponent has a minimum as the threshold value is varied.
 We wish to emphasize that dependence of on- and off-time exponents on threshold can qualitatively change from one trajectory to another.

We interpret this strong dependence on threshold as a clear indication that
the standard trajectory analysis - based on the separation between on and off events with a somewhat arbitrary threshold - is not quite adequate,
and the trajectories should be analyzed over the full range of threshold parameter.
 It also explains wide distribution of the exponents found by different groups.
As shown below, one of the key results of this paper is to exploit the threshold dependence of the trajectory parameters to retrieve important information
about the physical mechanism of the fluctuations.

The power spectrum of the fluorescence trajectory of a single QD
has a power law form \cite{PeltonAPL04,PeltonPNAS07}
$ S_I(\omega)\sim \omega^{-l} $ where $l$ is close to $1$. Therefore, we can consider the QD blinking
process as an example of single particle 1/f (flicker) noise. The
generally accepted phenomenological model for the electrical 1/f noise generation in
solids is that of electrical transport in the presence of an environment consisting of multiple stochastic two-level systems (TLS)\cite{Kogan,Weissman}. In the case of  QD blinking we suggest a similar physical model based on a TLS environment \cite{Grigolinicomment}.

 In our model the non-radiative relaxation of the QD excitation occurs via
trapping of holes to one of the $N$ quenching centers, followed by a non-radiative recombination with the remaining electron. Each of these  quenching centers could be dynamically switched between inactive and active conformations. The two conformational states differ in their ability to trap holes: the hole trapping rate is much larger in the active conformation than it is in the inactive state. Recent studies of trapping rates in the single QDs \cite{ScholesPNAS09} showed that the number of hole traps on the surface and  on the core/shell interface is in order of 10. Interestingly, we find that we  only need a similar number of recombination centers in order to reproduce the basic features of the fluorescence trajectories.
A possible microscopic origin of the conformation change in the recombination center could be due to the light-induced jumps of the surface or interface atom between two quasi-stable positions.
 The surface atoms in such  a small object as colloidal QD  can be found in a variety of local crystal configurations. Consequently, we can expect a wide distribution of switching rates.
 The non-radiative  trapping rate in our model can therefore be expressed as
\begin{equation}
k_t(t)=\sum_{i=1}^N k_i\sigma_i(t)+k_0.
\label{eq:TLS}
\end{equation}
For each TLS the stochastic variable $\sigma_i(t)$ randomly jumps between two
values $0$ and $1$, corresponding to inactive and active
conformations, respectively. Furthermore,  $k_i$ is the trapping rate in
the active configuration, and $k_0$ is the background non-radiative relaxation rate.
The time distribution functions for the $\sigma_i=0 \to 1$
transitions and $\sigma_i=1 \to 0$ transitions for the i-th TLS are
exponential and  can be characterized by the transition rates
$\gamma^+_i$ and $\gamma^-_i$, respectively. While in the simplest
model the transition rates for the individual TLS are constants
(non-interacting TLSs), we will show that a more general case of the
interacting TLS systems must also be considered.
  The power spectral density of the process (\ref{eq:TLS}) within the non-interacting TLS model is a sum of Lorentzians
\begin{equation}
S_k(\omega)=\frac 1 { \pi} \sum_{i=1}^N \frac {\gamma_i^+\gamma_i^-}{\gamma_i^++\gamma_i^-}\frac{k_i^2}{\omega^2+(\gamma_i^++\gamma_i^-)^2}.
\label{eq:Spec}
\end{equation}
The number of parameters in the above expression can be drastically reduced if the experimental constraint of $1/f$ noise spectrum is imposed. Indeed, after choosing $k_i=k$  and $\gamma^+_i=\gamma^-_i=\gamma_i\equiv\gamma_0 a^i$, where $a\ll 1$, one can effectively fit the spectrum in Eq. (\ref{eq:Spec}) with $1/\omega$ in the frequency region $\gamma_N \ll \omega \ll \gamma_1 $ \cite{Kogan}. Assuming low excitation intensities and  steady-state conditions for the fermionic degrees of freedom, the quantum yield $Y(t)$ is given by\cite{FrantsuzovPRB05}:
\begin{equation}
Y(t)=\frac {k_r} {k_r+k_t(t)}, \label{eq:Y}
\end{equation}
where  $k_r$ is the radiative relaxation rate.

Let us now show that our suggested model of fluorescence
fluctuations exhibits strong threshold dependence of the on- and
off-time distribution parameters. The problem of finding these distributions
for the stochastic process $Y(t)$ with known properties and threshold
value $y$  is equivalent to the well-known crossing problem
\cite{Stratonovich}. There are only few
cases when this problem can be solved exactly \cite{MajumdarCS99}.
Fortunately, our present model can be reduced to such an exactly solvable case.
The system at any moment $t$ could be completely described by the
configuration $\Sigma=\{\sigma_1,\dots,\sigma_N\}$. Clearly, there
are $2^N$ different configurations. A random walk in the given
configuration space is a Markovian stochastic process.  The vector
$\vec P$ containing probabilities of all configurations $P_\Sigma$ satisfies  the
Master equation
\begin{equation}
\frac d {dt} \vec P(t)=\hat W \vec P(t)
\label{eq:Master}
\end{equation}
where the transition matrix $\hat W$ contains the following nonzero
elements
$$W_{\Sigma^+_i\Sigma^-_i}=\gamma^+_i,\quad W_{\Sigma^-_i\Sigma^+_i}=\gamma^-_i, \quad W_{\Sigma\Sigma}=-\sum_{\Sigma'\neq \Sigma}W_{\Sigma'\Sigma},$$
where $\Sigma^-_i=\{\sigma_1,\dots,\sigma_i=0,\dots,\sigma_N\}$ and
 $\Sigma^+_i=\{\sigma_1,\dots,\sigma_i=1,\dots,\sigma_N\}$ for each given $\Sigma$.
The non-radiative relaxation rate for given configuration $\Sigma$
can be expressed by Eq. (\ref{eq:TLS}), which allows us to find the
corresponding emission intensity level $Y_\Sigma$ from Eq.
(\ref{eq:Y}). Let us introduce a threshold value  for the quantum
yield $y$. By definition, the QD is in the bright  (on) state if
$ Y(t)\ge y$ and dark  (off) state otherwise. For each threshold level $y$ all configurations  can be separated to a bright group,  satisfying a condition $Y_\Sigma \ge y$  and a dark group.  The vector of probabilities can be presented in the form
$\vec P=\left(\begin{array}{l} \vec P_b\\ \vec P_d \end{array} \right)$,
where vectors $\vec P_b$ and $\vec P_d$ contain probabilities of bright and
dark configurations, respectively.
The transition matrix can be recast in block form
$
\hat W=\left(\begin{array}{ll}
\hat W_{bb} & \hat W_{bd}\\
\hat W_{db} & \hat W_{dd}\\
\end{array}\right)$.
The expressions for the normalized on-time and off-time distribution functions in this
notations are well-known \cite{RiceJAP86}
\begin{eqnarray}
p_{\mbox{on}}(t)=\left\langle\vec 1, \hat W_{db} \exp(\hat W_{bb}t) \hat W_{bd} \vec P_{e}\right\rangle
\left\langle \vec 1,\hat W_{bd} \vec P_{e} \right\rangle^{-1} \nonumber\\
p_{\mbox{off}}(t)=\left\langle \vec 1, \hat W_{bd} \exp(\hat W_{dd}t) \hat W_{db} \vec P_{e} \right\rangle
\left\langle \vec 1, \hat W_{db} \vec P_{e} \right\rangle^{-1}
\label{eq:pon}
\end{eqnarray}
where $\langle \vec a,\vec b\rangle$   denotes the scalar product, $\vec 1$ is the unity vector
 and  $\vec P_{e}$ is  the equilibrium
probabilities vector, satisfying a stationary condition
$ \hat W\vec P_{e}=0 $.
We found that the on-time and off-time distributions generated
by Eqs. (\ref{eq:pon})  can be fitted by
a power law function (\ref{eq:PowerLaw}) (see the insert in Fig. 2a). Beyond a certain off/on time value  the power law behavior sharply
changes to exponential asymptotic behavior $\exp (-t/T)$. In our analysis, this value of $T$ is defined as a truncation time. We performed simulations of on-time and off-time distributions  for the model system of non-interacting TLS. This relatively simple model reproduces the general trend seen experimentally in the truncation times: The on-time truncation
decreases and the off-time truncation increases  when the threshold value
goes up.

While this simple, non-interacting TLS model is useful in illustrating our procedure, it cannot reproduce the threshold  dependence of the exponents.
 The slope of the on-time distribution monotonically increases with the threshold value, when the off-time exponent has
an opposite trend. In order to make our model more realistic, we introduce interaction between TLS in the
simplest mean-field form (similar to Ref.\cite{GrigoliniPA08}). The interaction is characterized by the parameter $\alpha$,
whereas the bias for
an individual TLS  is parameterized by $\beta$:
\begin{equation}
\gamma^{\pm}_i=\gamma_i\exp\left(\pm\,\alpha \sum_{i=1}^N (\sigma_i-1/2)\pm\beta\right)
\label{eq:Inter}
\end{equation}
 Fig. 2 provides the  numerical calculation results for the interacting model with the following parameters:
 $N=10$, $\gamma_1=1$, $a=10^{-1/2}$, $k_r/k=1$,  $k_0=0$,  $\alpha=0.27$ and
 $\beta=-0.13$.
As seen from this figure, the threshold dependence of the truncation times keep the same trend as for noninteracting case. In contrast, the slopes now show a non-monotonic
threshold dependence reproducing qualitatively the experimental behavior shown in Fig.1a. The insert in Fig. 2b shows that the interacting TLS model is capable of generating the two-maximum intensity distribution seen in Fig 1b. The relative ease with which our simple phenomenological model captured the experimental trend gives us hope that the model can be used to extract interaction parameters for the TLS environment. These parameters could provide useful experimental constraints on future {\it microscopic} models for the TLS environment of a variety of systems showing fluorescence intermittency.
The model proposed here also explains recent observations of the non-blinking dots. Furthermore, a similar model can be constructed for the fluorescence intermittency seen in quantum wires. The details for these results will be published elsewhere.
\begin{figure}[htp]
\includegraphics[width=3 in,clip]{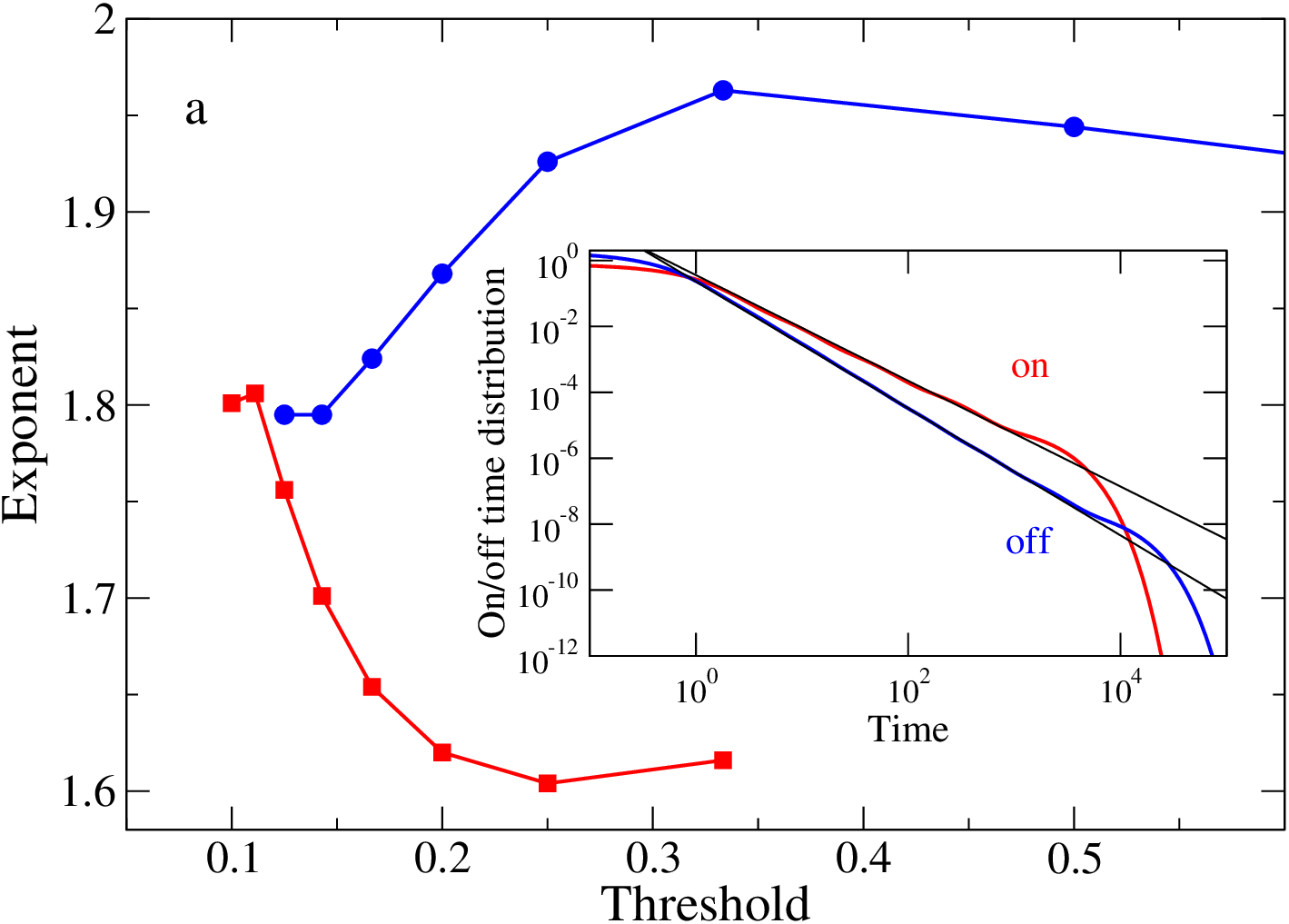}
\hfill
\includegraphics[width=3 in,clip]{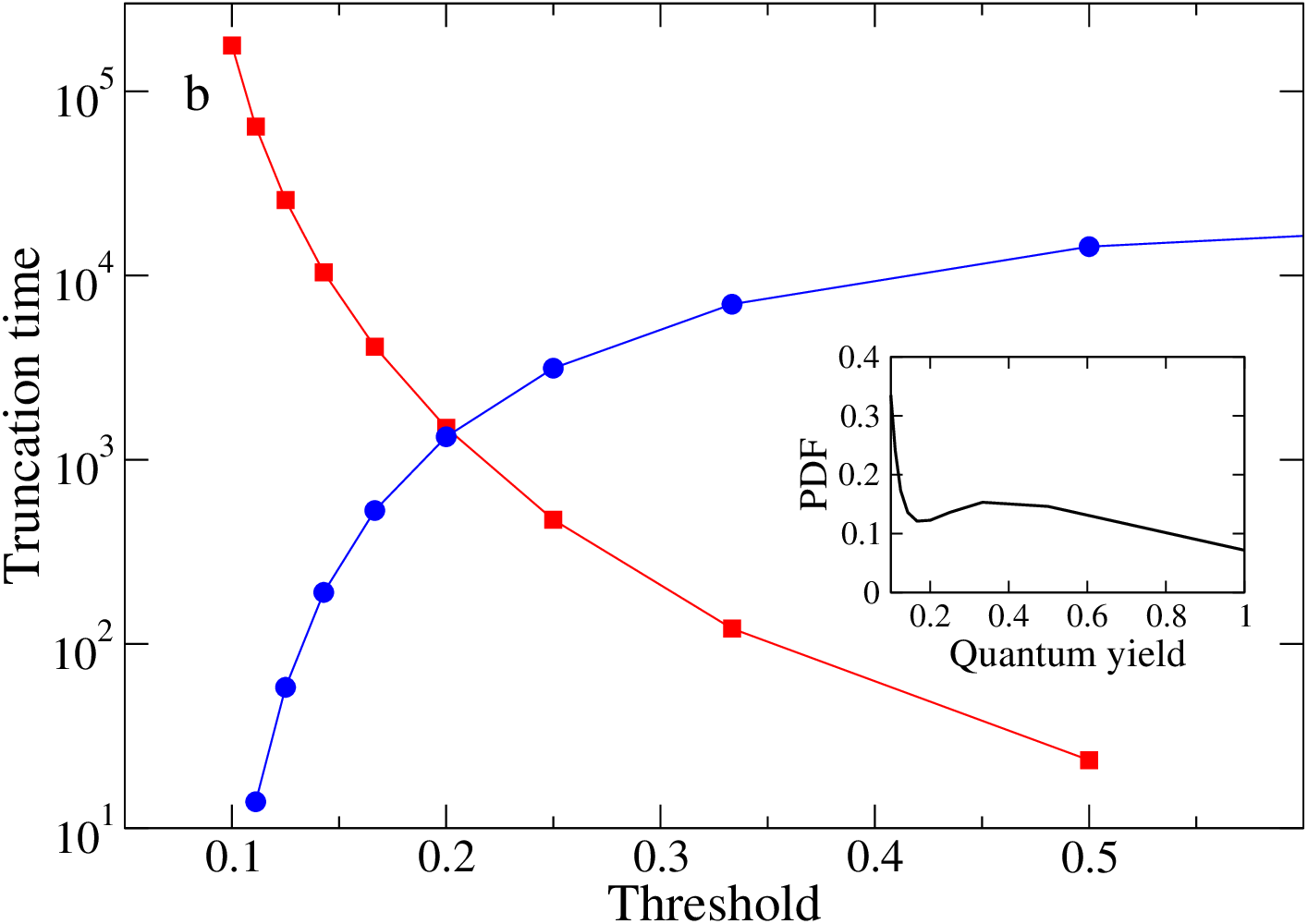}
\caption{(color online) The theoretical threshold dependence of the on-time (red squares) and off-time (blue circles) distribution exponents  (a) and  truncation times (b)
for the interacting TLS model (\ref{eq:Inter}). Insert a: the on- and off-time distribution functions at the threshold value y=0.25. Insert b:
probability distribution function (PDF) of the quantum yield.}
\end{figure}

In conclusion, the phenomenological model we proposed in this paper succeeds in qualitatively explaining the key experimental facts characterizing long-correlated fluorescence intensity fluctuations of the single colloidal quantum dots:
(1) the truncated power-law distributions for on- and off-times obtained by the commonly used threshold procedure;
(2) the strong threshold dependence of the distribution parameters $m$ and $T$ and wide range of the the extracted exponents;
(3) the 1/f noise form of the power spectrum of the intensity fluctuations;
(4) the continuous distribution of emission intensities and excitation lifetimes;
(5) the weak temperature dependence of the fluorescence intermittency due to the light-driven character of the TLS switching process.

We would like to thank Dr. Vladimir Protashenko and especially
Professor Masaru Kuno for many useful conversations and for providing us with high quality
experimental data. We would also like to acknowledge the support of
the Institute for Theoretical Sciences, the Department of Energy,
Basic Energy Sciences, and the National Science Foundation via the
NSF-NIRT grant No. ECS-0609249.

\end{document}